\documentstyle[epsfig,tables]{qcdparis1}
\def\beq{\begin{equation}}
\def\eeq{\end{equation}}
\def\bea{\begin{eqnarray}}
\def\eea{\end{eqnarray}}
\def\defeq{\equiv}
\newcommand{\qL}{q_{\scriptscriptstyle L}}
\newcommand{\qR}{q_{\scriptscriptstyle R}}
\newcommand{\qbL}{\bar{q}_{\scriptscriptstyle L}}
\newcommand{\qbR}{\bar{q}_{\scriptscriptstyle R}}

\newcommand{\syst}{~{\rm (syst.)}~}
\newcommand{\stat}{~{\rm (stat.)}~}
\newcommand{\naive}{na\"{\i}ve}
\newcommand{\role}{r\^ole}
\newcommand{\eqref}[1]{(\ref{#1})}   
\def\Toprel#1\over#2{\mathrel{\mathop{#2}\limits^{#1}}}

\def\etal{{\em et al.}}
\def\PL{{\em Phys. Lett.\ }}
\def\NP{{\em Nucl. Phys.\ }}
\def\PR{{\em Phys. Rev.\ }}
\def\PRL{{\em Phys. Rev. Lett.\ }}

\begin{document}
\pagestyle{plain}
\title{
\begin{flushright}
{\baselineskip12pt \normalsize
CERN-TH/95-279\\
TAUP-2297-95\\
hep-ph/9510402 \\
$\phantom{a}$
}
\end{flushright}
Nucleon Spin}

\author{John Ellis$^{\dag}$ and Marek Karliner$^{\ddag}$}

\affil{
$^{\dag}$Theory Division, CERN, CH-1211, Geneva 23, Switzerland. \\
e-mail: johne@cernvm.cern.ch \\
and\\
$^{\ddag}$School of Physics and Astronomy,
Raymond and Beverly Sackler Faculty of Exact Sciences
\\ Tel-Aviv University, 69978 Tel-Aviv, Israel
\\ e-mail: marek@vm.tau.ac.il}

\abstract{
We review the theory of polarized deep inelastic scattering
in light of the most recent experimental data.
We discuss
the nucleon spin decomposition and the Bjorken sum rule.
The latter is used for
extraction of $\alpha_s(M_Z^2)=0.116^{{+}0.004}_{{-}0.006}$
 \ and as a test case for a new method of analyzing
divergent perturbation series in QCD.
}
\resume{$\phantom{a}$}

\twocolumn[\maketitle]
\fnm{7}{
Based on plenary
talks presented by M. Karliner at the
Workshop on Deep Inelastic scattering and QCD,
Paris, April 1995 and at the
Workshop on the Prospects of Spin Physics at HERA,
DESY-Zeuthen, August 28-31, 1995.
}

\section{Analysis of Polarized Structure Functions}

\subsection{Formalism}

    The basis for our discussion will be the two spin-dependent structure
functions $G_1$ and $G_2$:
\bea
\frac{d^2\sigma^{\uparrow\downarrow}}{dQ^2d\nu} -
\frac{d^2\sigma^{\uparrow\uparrow}}
{dQ^2d\nu} =
\phantom{aaaaaaaaaaaaaaaaaaaaaaaaaaa}
\nonumber \\
\frac{4\pi\alpha^2}{Q^2E^2}~\bigg[M_N(E+E^{\prime}\cos
\theta )G_1(\nu ,Q^2)
- Q^2G_2(\nu ,Q^2)\bigg]
\label{E1}
\eea
In the parton model, these structure functions scale as follows in the
Bjorken limit
$x = {Q^2}/{2M_N\nu}$ fixed,  $Q^2\rightarrow\infty$:
\hfill\break
\vbox{
\bea
M_N^2\nu\, G_1(\nu ,Q^2) \defeq g_1(x,Q^2)
\rightarrow g_1(x) \nonumber\\
\label{E3} \\
M_N\nu^2
G_2(\nu ,Q^2) \defeq g_2(x,Q^2) \rightarrow g_2(x)
\nonumber
\eea
}
\hfill\break
We will discuss the scaling structure function $g_2$ later on, focussing
for now on $g_1$, which is related to the polarized quark distributions
by
\bea
g^p_1(x) &=& {1\over 2} \sum_q~e^2_q[q_{\uparrow}(x) - q_{\downarrow}(x)
+ \bar
q_{\uparrow}(x) - \bar q_{\downarrow}(x)]\\ \nonumber
&=&  {1\over 2} \sum_q~\Delta q(x)
\label{E4}
\eea
for comparison, the unpolarized structure function $F_2$ is given by
\beq
F_2(x) = \sum_q e^2_qx[q_{\uparrow}(x) + q_{\downarrow}(x) + \bar
q_{\uparrow}(x) -
\bar q_{\downarrow}(x)]
\label{E5}
\eeq
so that the polarization asymmetry $A_1$ may be written as
\beq
A_1 = \frac{\sigma_{1/2}-\sigma_{3/2}}{\sigma_{1/2} + \sigma_{3/2}}
\label{E6}
\eeq
in the Bjorken limit, where $\sigma_{1/2}$ and
$\sigma_{3/2}$ are the virtual
photon absorption cross sections.  We will discuss later the
$Q^2$ dependences of the above formulae, as well as the transverse
polarization asymmetry.

    Much of the interest in the polarized structure function $g_1$ is due
to its relation to axial current matrix elements:
\bea
\langle p\vert A^q_{\mu}\vert p\rangle = \langle p\vert\bar
q\gamma_{\mu}\gamma_5q\vert
p\rangle =
\langle p\vert\qbR \gamma_{\mu}\qR -
\qbL\gamma_{\mu}\qL\vert
p\rangle =
\nonumber\\
=\Delta q\cdot S_{\mu}(p)
\phantom{aaaaaaaaaaaaaaaaaaaaaaaaa}
\nonumber
\label{E8}\\
\eea
where $q_{\scriptscriptstyle L,R}
\equiv 1/2 (1 \mp \gamma_5) q$, $S_{\mu}$ is the nucleon
spin
four-vector, and
\beq
\Delta q \equiv \int^1_0 dx[q_{\uparrow}(x) - q_{\downarrow}(x) + \bar
q_{\uparrow}(x) -
\bar q_{\downarrow}(x)]
\label{E9}
\eeq
Of particular interest is the matrix element of the singlet axial
current
\beq
A^0_{\mu} = \sum_{q=u,d,s} \bar q\gamma_{\mu}\gamma_5q:\quad\quad
\langle p\vert A^0_{\mu}\vert p\rangle =
\sum_{q=u,d,s} \Delta q\cdot S_{\mu}(p)
\label{E10}
\eeq
which is related in the parton model to the sum of the light quark
contributions to the proton spin.  Prior to the series of measurements
of polarized deep inelastic lepton nuclear scattering, information
was available from charged current weak interactions on some axial
current matrix elements.  For example, neutron beta decay and strong
isospin symmetry tell us that \cite{RPP}
\beq
\Delta u - \Delta d = F+D = 1.2573 \pm 0.0028
\label{E12}
\eeq
and hyperon beta decays and flavour $SU(3)$ symmetry tell us that
\cite{FoverD}
\beq
\frac{\Delta u + \Delta u - 2\Delta s}{\sqrt{3}}
\equiv {a_8 \over \sqrt{3}}
= {3F -D \over \sqrt{3}} = 0.34 \pm 0.02
\label{E13}
\eeq
(for a recent discussion of the applicability of $SU(3)$
symmetry see Ref.~\cite{LL} and references therein).
Equations \eqref{E12} and \eqref{E13}
give us two equations for the three unknowns $\Delta u$, $\Delta d$
and $\Delta s$.  In principle, a third piece of information was available
\cite{EK,KaplanManohar}
in 1987 from neutral current weak interactions.  Measurements of
$\nu p$ and $\bar \nu p$ elastic scattering \cite{Ahrens} indicated that
\beq
\Delta s = -0.15 \pm 0.09
\label{E15}
\eeq
but this information was not generally appreciated before the advent
of the EMC data discussed below. At present there is a new
neutrino experiment under way at Los-Alamos \cite{Garvey}
which is expected to significantly improve the precision of
\eqref{E15} (see Ref.~\cite{Alberico} for a recent in-depth analysis).

    In the \naive\ parton model, the integrals of the $g_1$ structure
functions for the proton and neutron
\bea
\Gamma_1^p(Q^2) \equiv \int^1_0 dx~g^p_1(x,Q^2)
\nonumber\\
\label{GammaIdefs}\\
\Gamma_1^n(Q^2) \equiv \int^1_0 dx~g^n_1(x,Q^2)
\nonumber
\eea
are related to combinations of the $\Delta q$.
\bea
\Gamma_1^p = {1\over2}\left(
{4\over9}\Delta u
+{1\over9}\Delta d
+{1\over9}\Delta s\right)
\nonumber\\
\label{GammaIDq}\\
\Gamma_1^p = {1\over2}\left(
{1\over9}\Delta u
+{4\over9}\Delta d
+{1\over9}\Delta s\right)
\nonumber
\eea

The difference between the proton and neutron integrals yields the
celebrated Bjorken sum rule \cite{BJ}
\beq
\Gamma_1^p(Q^2) - \Gamma_1^n(Q^2) =
{1\over 6}
(\Delta u-\Delta
d)\times (1-\alpha_s(Q^2)/\pi ) + \,\,\ldots
\label{E16}
\eeq

It is not possible to derive individual sum rules for $\Gamma_1^{p,n}$
without supplementary assumptions.  The assumption made
by Ellis and Jaffe in 1973 \cite{EJ}
was that $\Delta s = 0$, on the grounds that very possibly there were
a negligible number of strange quarks in the nucleon wave function,
and if there were, surely they would not be polarized.  With this
assumption, it was estimated that
\bea
\int_0^1 dx g^p_1 (x,Q^2)
{=}{1\over 18} (4\Delta u + \Delta d)~(1-\alpha_s/\pi + \ldots) =
\nonumber\\
\label{E17}\\
= 0.17 \pm 0.01\,
\phantom{aaaaaaaaaaaaaaaaaaa}
\nonumber
\eea
It should be clear that this was never a rigorous prediction, and was
only intended as a qualitative indication to experimentalists of what
they might find when they started to do polarized electron proton
scattering experiments.

    Perturbative QCD corrections to the above relations have been
calculated \cite{Kodaira}-\cite{Larin}:
\bea
\int^1_0 [ \,g_1^p(x,Q^2) - g_1^n(x,Q^2)\,] =
{1 \over 6}\, |g_A| \, f(x)\,\,:
\nonumber\\
\label{bjf}\\
f(x) = 1 - x - 3.58 x^2 - 20.22 x^3 + \,\ldots
\nonumber
\eea
and
\hfill\break
\vbox{
\bea
\int^1_0 \,g_1^{p(n)}(x,Q^2) =
\phantom{aaaaaaaaaaaaaaaaaaaaaaaaa}
\nonumber\\
=\left(\pm{1\over12}\,|g_A| +{1\over36}\,a_8\right)\,f(x)
+{1\over9}\,\Delta\Sigma(Q^2)\,h(x)\,\,:
\label{ejCorr}\\
h(x) = 1 - x - 1.096 x^2 - \,\ldots
\phantom{aaaaaaaaaaaaaaaa}
\nonumber
\eea
}
where $x = \alpha_s(Q^2)/\pi$, and the dots
represent uncalculated higher orders of perturbation theory,
to which must be added higher-twist corrections which we will discuss
later.
The coefficients in \eqref{bjf},\eqref{ejCorr}
are for $N_f{=}3$, as relevant for
the $Q^2$ range of current experiments.

With these corrections, the Bjorken sum rule is a
fundamental prediction of QCD which can be used, for example, to
estimate a value for $\alpha_s(Q^2)$.  On the other hand, the individual
proton and neutron integrals can be used to extract a value of
$\Delta s$.

\subsection{The Helen of spin}

    Early data on polarized electron-proton scattering from SLAC-Yale
experiments \cite{oldSLACa,oldSLACb,oldSLACc}
were compatible with the prediction of equation
\eqref{E17} within
large errors.  Over a 1000 theoretical and experimental papers
were launched by the 1987 EMC result \cite{EMC}
\bea
\int^1_0 g^p_1(x,Q^2)=0.126 \pm 0.010 \syst\pm 0.015\stat\ , \quad
\kern-1cm\phantom{a}
\nonumber\\
\hbox{at}~\langle Q^2 \rangle = 10.7~{\rm GeV}^2
\phantom{aaaaaaaaaaaaaaaaaa}
\label{E19}
\eea
which was in {\em prima facie} disagreement with the dynamical assumption
that \hbox{$\Delta s = 0$}.
 It is worth pointing out that the small-$x$ behaviour
of $g_1^p(x)$ was crucial to this conclusion.  The earlier SLAC-Yale
data had large extrapolation errors, and the EMC data indicated behaviour
different from that in simple dynamical models.  They were, however,
consistent \cite{EK}
with the \naive\ Regge expectation \cite{Heimann}
\beq
g^p_1(x) \simeq \sum_i c_i\,x^{-\alpha_i(0)}
\label{E20}
\eeq
were the $\alpha_i(0)$ are
the intercepts of axial vector Regge trajectories
which are expected to lie between 0 and -0.5.   A fit to the
EMC data gave \cite{EK}
\beq
g^p_1 \sim x^{-\delta} : \quad \delta = -0.07^{+0.42}_{-0.32}~~{\rm
for}~~x < 0.2
\label{E21}
\eeq
for $x<0.2$.

    Using equations \eqref{E12},\eqref{E13},\eqref{E19}
 and the leading order perturbative QCD corrections
in equation \eqref{ejCorr} it was estimated \cite{BEK} that
\hfill\break
\medskip
\vbox{
\bea
\Delta u &=& \phantom{-}0.78 \pm 0.06 \nonumber \\
\Delta d &=& -0.47 \pm 0.06 \label{E23} \\
\Delta s &=& -0.19 \pm 0.06 \nonumber
\eea
}
\medskip
Strikingly, these determinations corresponded to a total contribution
of quarks to the proton spin
\beq
\Delta\Sigma =
\Delta u + \Delta d + \Delta s = 0.12 \pm 0.17
\label{E24}
\eeq
which was compatible with 0.   This has sometimes been called the
``proton spin crisis", but we think this is an over-reaction.  The
result equation \eqref{E24} was certainly a surprise for \naive\ models of non-
perturbative QCD, but it was not in conflict with perturbative QCD.
Moreover, shortly after the first data
became available it was shown \cite{BEK} that $\Delta\Sigma = 0 $
occurs naturally in the Skyrme model, which is
believed to reproduce the essential features of QCD in
the large-$N_c$ limit.
Alternatively, it was suggested
that the $U(1)$ axial anomaly
and polarized glue might provide
an alternative interpretation \hbox{\cite{DeltagI}-\cite{DeltagIII}},
or a significant suppression of the QCD topological
susceptibility \cite{SVa}-\cite{NSV}
might play a key \role, which would modify the \naive\ quark
model predictions.

In 1972 Richard Feynman wrote
``$\dots$ {\em its [the Bjorken sum rule's] verification,
or failure, would have a most decisive effect on the direction of
future high-energy physics}".  On the other hand, we think that the
verification, or failure, of equation \eqref{E24}
has only an indecisive effect, though a very interesting one.

\subsection{Evaluation of integrals}

\begin{figure}[htb]
\begin{center}
\mbox{\epsfig{file=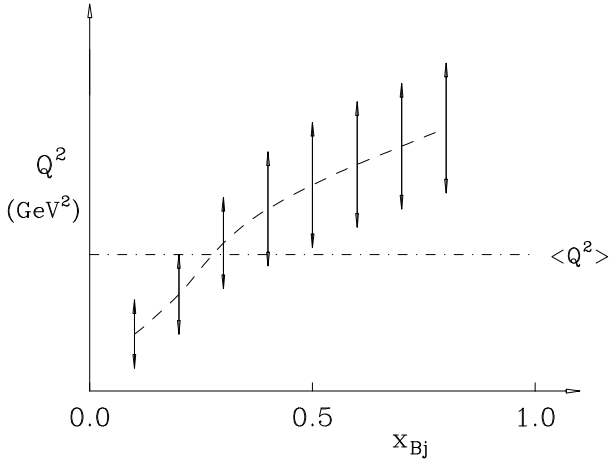,width=8truecm,angle=90}}
\caption{
In \ any \ given \ polarized \ lepton-nucleon \ scattering
expe\-riment,
the range of \protect $Q^2$ probed is different in different bins of
the Bjorken variable \protect  $x_{B_j}$.
}
\label{FigI}
\end{center}
\end{figure}

    Before discussing the interpretation of more recent data on polarized
structure functions, we first review a few points that arise in the
evaluation of the integrals $\Gamma_1^{p,n}$.  It should not be forgotten
that the QCD versions of the sum rules are formulated at fixed $Q^2$.
 A generic deep-inelastic sum rule in QCD reads
\beq
\Gamma (Q^2) = \Gamma_{\infty} \bigg[ 1 + \sum_{n \ge 1}
c_n~\bigg({\alpha_s(Q^2)\over\pi}\bigg)^n\bigg] + \sum_{m \geq
1}~{d_m\over (Q^2)^m}
\label{E33}
\eeq
where $\Gamma_{\infty}$ is the asymptotic value of the sum rule for
$Q^2\rightarrow\infty$, the $c_n$ are the coefficients of the
perturbative corrections,
and the $d_m$ are  coefficients of the so-called mass and higher-twist
corrections.
On the other hand,
the data are normally obtained at values of $Q^2$ that increase
on the average with $x_{Bj}$ as seen in Fig.~\ref{FigI}.

 It is therefore necessary
for each individual experiment to interpolate and extrapolate to some
fixed mean value of $Q^2$, as indicated by the dashed horizonal line
in Fig.~\ref{FigI}.
The quantity measured directly is the polarization asymmetry
\eqref{E6},
which seems experimentally to have only small dependence on $Q^2$.
Therefore experiments often assume that $A_1$ is a function of $x$ only,
and then estimate
\bea
g_1(x,Q^2) = {A_1(x,Q^2)F_2(x,Q^2)\over 2x [1 + R(x,Q^2)]}
\simeq
{A_1(x)F_2(x,Q^2)\over 2x [1 + R(x,Q^2)]}
\nonumber\\
\label{E34}
\eea
where $F_2 (x,Q^2)$ and $R(x,Q^2)$
(the ratio of longitudinal to transverse virtual photon cross-sections)
are taken from parametrizations of
unpolarized scattering data.  Note that these induce a $Q^2$ dependence
in $g_1$ even if $A_1$ is independent of $Q^2$.

    The possible reliability of the assumption that $A_1$ is independent
of $Q^2$ can be explored using leading-order perturbative QCD models for
$g_1(x,Q^2)$.  Several such studies have been
made \cite{ANR}-\cite{GluckPol},
and they indicate
a small $Q^2$ dependence in $A_1 (x,Q^2)$.
This has in turn a small effect
on the extracted values of $g_1(x,Q^2)$, which is much smaller than the
statistical errors of the EMC and SMC experiments and is not very
significant for the E142 and E143 experiments, and is in any case
considerably smaller than the systematic errors, so that it does not
yet contribute an important error to the evaluations of the
$\Gamma_1^{p,n}$.  However, it could become an important effect in the
future, and both theorists and experimentalists should keep their
eyes open.

    The old-fashioned assumption of Regge behaviour at low x also needs
to be checked carefully.  The leading-order perturbative QCD evolution
equations for the non-singlet part of the helicity distributions,
$\Delta q_{NS} (x,Q^2)$, lead us to expect
\cite{DGPTWZ,BEKlowx}
singular behaviour
as $x \rightarrow 0$, so that
\beq
\Delta q_{NS}(x,Q^2) \simeq
C_{NS} \exp(A_{NS}\sigma+ B_{NS} {\sigma\over\rho}
-\ln \rho -{1\over2} \ln \sigma)
\label{DqNS}
\eeq
where $A_{NS}$, $ B_{NS}$ and $C_{NS}$ are some constants
and
\beq
\sigma \equiv \sqrt{\ln{x_0\over x}\ln {t\over t_0}}\,, \quad
\rho \equiv \sqrt{{\ln {x_0\over x}\over {\ln {t\over t_0}}}}\,,
\quad
t \equiv \left(\ln{Q^2\over\Lambda^2}\over\ln{Q^2_0\over\Lambda^2}\right)
\label{DqNSdefs}
\eeq
and we might expect by analogy with the unpolarized structure
functions that $x_0 \sim 0.1$,  $Q^2_0 \sim 1$ GeV and the
leading-order QCD scale parameter $\Lambda \sim 0.25$ GeV, with
\beq
A_{NS} = {4\sqrt 2\over \sqrt{33-2N_f}}\,, \quad B_{NS} = {4\over
33-2N_f} \ .
\label{DqNSdefsII}
\eeq
where $N_f=3$ in the $Q^2$ range of current experimental
interest (see also Ref.~\cite{BallFortePol}).
In principle
Eq.~\eqref{DqNS}
can be applied directly to the low-$x$ behavior of the
integrand of the Bjorken sum rule:
$g^p_1(x,Q^2)-g^n_1(x,Q^2)={1\over6}[\Delta u(x,Q^2)-\Delta d(x,Q^2)]$,
as well as to the other nonsinglet combination $\Delta u(x,Q^2)+
\Delta d(x,Q^2)-2\Delta s(x,Q^2)$ that also contributes to
$g_1^{p,n}(x,Q^2)$.
Note, however, that effects similar those in the BFKL pomeron may also be
important at very low $x$ \cite{BER}.
The flavour-singlet
combination of structure functions has a more complicated low-$x$
behaviour, which could be important for the extraction of the
$\Delta q$.
More singular low-$x$ behaviours have been proposed in the literature
motivated by non-perturbative QCD considerations
\hbox{\cite{BassLandshoff,CloseRobertsG1}}.
  It is not clear whether the behaviour in equation \eqref{DqNS}
is relevant to the data presently available:  one SMC data point may
be in its region of applicability, and could in principle be used
to normalize the perturbative QCD formula, serving as a basis for
extrapolating the integrals to $x=0$.  In practice, it does not seem
at present that this would have a significant effect on the evaluation
of the Bjorken sum rule.

    The analysis of the polarized structure function data has often
assumed that the transverse polarization asymmetry
\beq
  A_\perp
= \frac{d \sigma^{\downarrow\rightarrow} -
              d \sigma^{\uparrow\rightarrow} }
             {d \sigma^{\downarrow\rightarrow} +
              d \sigma^{\uparrow\rightarrow} }\;.
\label{Atransverse}
\eeq
is negligible.  This is related to the spin-flip photon absorption
asymmetry
\beq
 A_2 = \frac{2\sigma_{TL}}{\sigma_{1/2}+\sigma_{3/2}}
\label{A2def}
\eeq
and the longitudinal $A_1$ asymmetry \eqref{E6}
through the relation:
\beq
A_\perp = \, d \ \left(A_2-\gamma\,(1-\frac{y}{2})\,A_1\right)\;.
\label{AperpA2}
\eeq
were $\sigma_{1/2}$ and $\sigma_{3/2}$ are the virtual photon--nucleon
 absorption
 cross sections
for total helicity 1/2 and 3/2, respectively, $\sigma_{TL}$
arises from the helicity spin-flip amplitude in forward photon--nucleon
Compton scattering,
$\gamma=2 M x/\sqrt{Q^2}$,
and
$y=\nu/E_{lepton}$,
where $\nu$ is
the energy transfer in the laboratory frame.
The coefficient $d$ is related to the virtual photon depolarization
factor $D$ by
\beq
d = D\frac{\sqrt{1-y}}{1-y/2}
\label{dD}
\eeq
The asymmetries $A_1$ and $A_2$ are subject to the following positivity
conditions \cite{POS}
\beq
    |A_1|<1\;, \qquad |A_2| \leq \sqrt{R}\; .
\label{PosCond}
\eeq
and are related to the structure functions $g_{1,2}$ by
\begin{eqnarray}
    A_1 = \frac{1}{F_1}(g_1-\gamma^2g_2)\;, \hspace{1cm}
    A_2 = \frac{\gamma}{F_1}(g_1+g_2)\;,
\label{A2}
\end{eqnarray}
where
$F_1=F_2(1+\gamma^2)/2x(1+R)$ is the spin-independent
structure function.
Recently, data on $g_2$ have become available for the first time
\cite{SMCg2,E143g2}.
They indicate that it is considerably smaller than the positivity bound
in equation \eqref{PosCond}, and is very close to the leading twist formula
of ref.~\cite{WW}:
\begin{equation}
    g_{2}^{\rm ww}(x,Q^2) = - g_1(x,Q^2) +\int_{x}^{1} g_1(t,Q^2)
\frac{d t}{t}\;,
\label{g2ww}
\end{equation}
The data on $g_2$ are also compatible with the Burkhardt-Cottingham
sum rule \cite{BC}
\begin{equation}
    \int_{0}^{1} g_2(x,Q^2) d x = 0\;,
\label{BCSR}
\end{equation}
which has been verified to leading order in perturbative QCD.
\cite{ALNR,KodairaBC}. However,
the experimental errors are still considerable, in particular because
the low-x behaviour of $g_2$ is less well understood than that of $g_1$.
However, the data already tell us that the uncertainty in $g_2$ is not
significant for the evaluations of the $\Gamma_1^{p,n}$.

\subsection{Higher orders in QCD perturbation theory:}

    The perturbation series in QCD is expected to be asymptotic with
rapidly growing coefficients:
\beq
S(x) = \sum^\infty_{n=0} c_n x^n~, \quad x \equiv \frac{\alpha_s}{\pi}~,
c_n
\simeq n!K^n {n}^\gamma
\label{GenericSeries}
\eeq
for some coefficients $K, \gamma$ \cite{renmvz,RenormRev}.
This type of behaviour is associated with the presence of the
renormalon singularities, as we shall discuss shortly.  Such series
are often evaluated approximately by calculating up to the ``optimal"
order, implicitly defined by
\beq
| c_{n_{opt}} \, x^{n_{opt}} |
< | c_{n_{opt}+1} \, x^{n_{opt}+1} |
\label{noptDef}
\eeq
and assuming an error of the same order of magnitude as
$c_{n_{opt}}\,x^{n_{opt}}$.  The question arises whether one can approach
or even surpass this accuracy without calculating all the terms up
to order $n_{opt}$.  This possibility has been studied using the
effective charge (ECH) approach
\cite{EffectiveCharge,KataevStarshenko}
and using commensurate scale
relations \cite{BLM,CSR}.
 In this section, we discuss the use of Pad\'e approximants
(PA's) for this purpose \cite{SEK,PBB}.

\begin{figure}[htb]
\begin{center}
\mbox{\epsfig{file=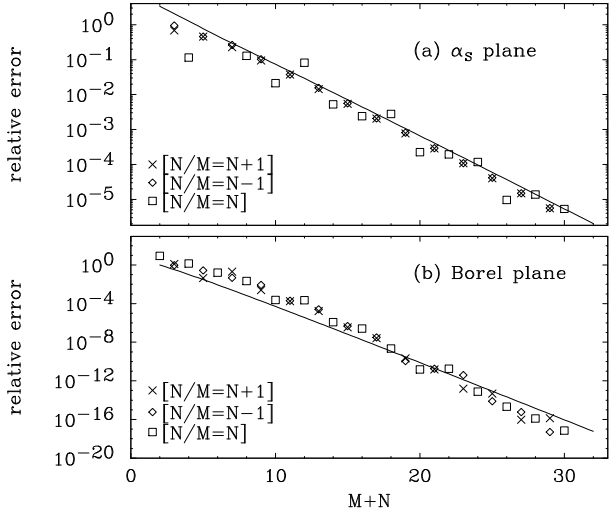,width=8.0truecm,angle=90}}
\end{center}
\caption{
\ Relative \ errors \ in \ the \ $[N/M]$ Pad\'e \ approximants
\hfill\break
(a) \ to \ the \
QCD \ vacuum \ polarization \ D-function, \ evaluated \ to \ all
\ orders
in \ the \ large-$N_f$ \ approximation
\protect\cite{Dfunction} \
(the rate of convergence agrees with expectations for
a series with \ a discrete set \ of Borel poles),
\ \ and \
\ (b) \ to \ the \ Borel \ transform \ of \ the D-function \ series,
where the convergence is particularly striking.
The straight lines correspond to the error formulae,
eqs.~\protect\eqref{Efive} and \protect\eqref{Eeight}, respectively.
\label{FigIII}
}
\end{figure}

    Pad\'e approximants \cite{Baker,BenderOrszag}
are rational functions chosen to equal the
perturbative series to the order calculated:
\bea
[N/M] = \frac{a_0 + a_1x + ... +a_Nx^N}{1 + b_1x + ... + b_Mx^M}~:
\nonumber\\
\label{PadeDef}\\
\lbrack N/M] = S + {\cal O}(x^{N+M+1})
\phantom{aaa}
\nonumber
\eea
Under certain circumstances, an expansion of the PA in equation
\eqref{PadeDef}
provides a good estimate,
$c^{est}_{N+M+1}$,
the Pade Approximant Prediction (PAP),
for the next coefficient $c_{N+M+1}$ in the perturbative series
\cite{PAPconvergence}.
  For
example, we have demonstrated that if
\beq
\epsilon_n \equiv \frac{c_{n}\, c_{n+2}}{c^2_{n+1}} - 1
\,\simeq \,{1\over n},
\label{epsDef}
\eeq
as is the case for any series dominated by a finite number of
renormalon singularities, then $\delta_{[N/M]}$ defined by
\beq
\delta_{[N/M]} \equiv \frac{c^{est.}_{N+M+1} -
c_{N+M+1}}{c_{N+M+1}}
\label{Efour}
\eeq
has the following asymptotic behaviour
\beq
\delta_{[N/M]} \simeq
-\,\frac{M!}{L^M  }\,, \quad~{\rm where}~ L = N+M+a\,M
\label{Efive}
\eeq
and where $a$ is a number of order 1 that depends on the series under
consideration.  This prediction agrees very well with the known
errors in the PAP's for the QCD vacuum polarization D function
calculated in the large $N_f$ approximation \cite{Dfunction},
as seen in Fig.~\ref{FigIII}a.

Large-$N_f$ calculations of the perturbative corrections to the
Bjorken sum rule \cite{largeNfBjSR}
indicate the presence of only a finite number of
renormalon singularities, so that the PAP's should be accurate.
Using the known terms in equation \eqref{bjf}, the [1/2] and [2/1] PAP's
yield the following estimates for
the fourth-order coefficient \cite{PBB}:
\bea
c^{Bj}_{4[PA]}\approx {-}111\qquad (\,\hbox{[1/2] \ PA)}
\nonumber\\
\label{bjpa}\\
c^{Bj}_{4[PA]}\approx {-}114\qquad (\,\hbox{[2/1] \ PA)}
\nonumber
\eea
and the error estimator in equation \eqref{Efive} with $a=1$ yields
\beq
\delta_{[1/2]} \simeq {-}1/8; \qquad
\delta_{[2/1]} \simeq {-}1/4
\label{bje}
\eeq
These results can be combined to obtain
\beq
c^{Bj}_{4[PA]}=
{1\over2}\left(\,
{-111\over 1 + \delta_{[1/2]}}
\,+\,
{-114\over 1 + \delta_{[2/1]}}
\,\right)
\approx  {-}139
\eeq
which is very close to the ECH estimate \cite{KataevStarshenko}
\beq
c^{Bj}_{4[ECH]} \simeq {-}130
\label{bjech}
\eeq

\begin{figure}[htb]
\begin{center}
\mbox{\epsfig{file=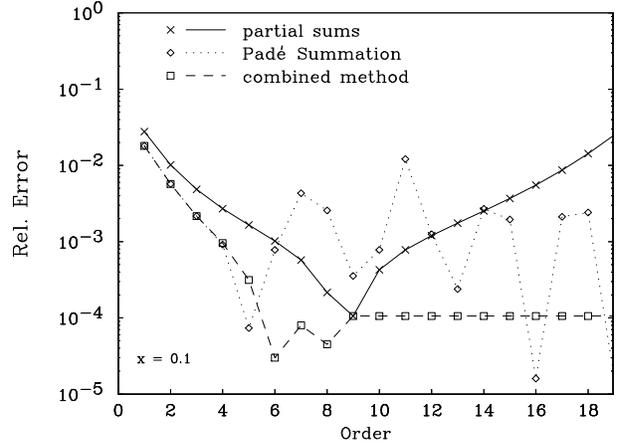,width=8truecm,angle=90}}
\end{center}
\caption{
The relative errors between partial sums of the series
$S(x) = \Sigma n! x^n$ and the Cauchy principal value of the series
(solid line)
is compared with the relative errors of Pad\'e \ Sums \ (dotted line).
\ We \ see \ that \ the \ relative
errors \ of \ the \ Pad\'e \ Sums are smaller
than those of the partial sums in
low orders, \ fluctuate in an intermediate r\'egime, \ and \ are again
\ more \ accurate than the partial sums \ in \ higher orders. \ The
fluctuations \ are associated with nearby poles in the Pad\'e Sums,
that may be treated by the ``combined method"
mentioned in the text, \ shown as the dashed line.
\label{FigIV}
}
\end{figure}

    A second application of PA's is to ``sum" the full perturbative
series.  The latter is ambiguous if the perturbative series possesses
an infrared renormalon singularity, i.e. a divergence of the form
in equation \eqref{GenericSeries} with $K > 0$.
Consider the following toy example:
\beq
\sum\limits_0^\infty n! \, x^n =
\int^\infty_0~{e^{-t}\over 1-xt}~~dt = {1\over x}
\int^\infty_0~{e^{-y/x}\over
1-y}~dy
\label{Eseven}
\eeq
which exhibits an infrared renormalon pole at $y = 1$.  One possible
way to define the ambiguous integral on the right hand side of
equation \eqref{Eseven} is via the Cauchy principle
value prescription \cite{PvaluePrescription}.
We see
in Fig.~\ref{FigIV}
that the errors in the Pad\'e "Sums" (PS's) $[N/M] (x)$ are
smaller than the truncated perturbative series
\beq
\sum\limits_{n=0}^{N+M} n! \, x^n
\eeq
when $n < n_{opt}$,
which is 5 in this example.  You will notice in
Fig.~\ref{FigIV} that the errors in the PS's become
unstable for large $n$:  this is
because of nearby poles in the denominator of equation
$\eqref{PadeDef}$ which are
not important for small $n$.  Also shown in Fig.~\ref{FigIV}
 as ``combined method"
is a systematic approach to treating these poles and optimizing the PS's
for large $n$, which is described elsewhere \cite{next}.

\begin{figure}[htb]
\begin{center}
\mbox{\epsfig{file=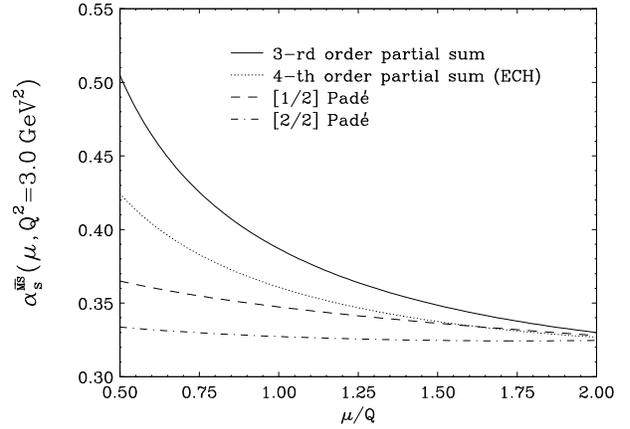,width=8truecm,angle=90}}
\end{center}
\caption{
\hyphenpenalty=-1000
The scale dependence of $\alpha_s(3 {\rm GeV}^2)$ obtained from
a \ fixed \ value
\ \ $f(x) = (6/ g_A) \times 0.164 =0.783$,
\ ({\em cf.} eq.~\protect\eqref{N}$\,$)\,,
\ for $Q/2 < \mu < 2Q$, \ \ using \ the \ \naive\ \ third- \ and
\ fourth-order
per\-turbative series and the [1/2] and [2/2] PS's.
\label{FigV}
}
\end{figure}

    Evidence that the PS's for the Bjorken sum rule provide a good
estimate of the perturbative correction factor in equation
\eqref{bjf} is
provided by the study of the renormalization scale dependence.  We see
in Fig.~\ref{FigV}
that the renormalization scale dependence of the [2/2]
PS is much smaller than that of the [2/1] and [1/2] PS's, which is
in turn much smaller than that of the \naive\ perturbation series
evaluated to third order.  We recall \cite{CollinsRG}
that the full correction factor
should be scale-independent, and interpret Fig.~\ref{FigV}
as indicating that the PS's may be very close to the true result.

\begin{figure}[htb]
\begin{center}
\mbox{\epsfig{file=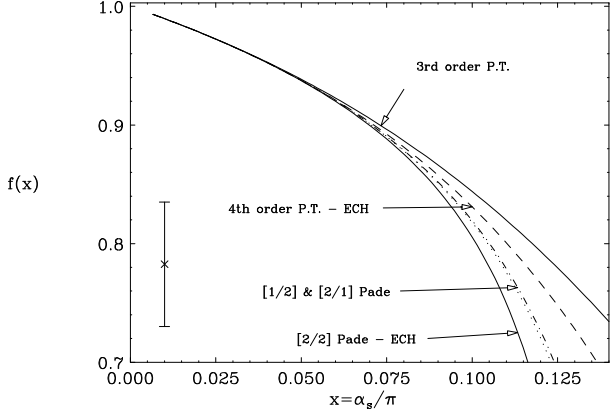,width=8truecm,angle=90}}
\end{center}
\caption{
\hyphenpenalty=-1000
Different \ approximations \ to \ the \ Bjorken \ sum \ rule correction
factor $f(x)$, third-order and fourth-order perturbation theory,
\ [1/2], \ [2/1] and \ [2/2] \ Pad\'e \ Sums
\ are compared. Also shown
as a vertical error bar is the value of $f(x)$
we extract from the available polarized structure data
\protect\eqref{N}.
\label{FigVI}
}
\end{figure}

    Fig.~\ref{FigVI} shows the estimates of the perturbative QCD correction
to the Bjorken sum rule obtained in various different approximations,
including third-order perturbation theory, fourth-order perturbation
theory estimated using the ECH technique and the [2/1], [1/2] and
[2/2] PS's.  We interpret the latter as the best estimator, and
take the difference between it and the [2/1] and [1/2] PS's as a
theoretical uncertainty.  Also shown in Fig.~\ref{FigVI} is the experimental
error on this quantity, as extracted from the combined analysis
of the available experimental data discussed in the next section.

    More information can be extracted by considering PA's in the
Borel plane.  The Borel transform of a perturbative series is
defined by
\bea
S(x) = \sum_{n=0}^\infty c_n x^n
\mathop{\quad\longrightarrow\quad}\limits_{\hbox{\scriptsize Borel}}
\phantom{aaaaaaaaaaaaa}
\nonumber\\
\tilde S(y) \equiv \sum^\infty_{n=0} \tilde c_n y^n~:~\tilde c_n =
\frac{c_{n+1}}{n!}\,\left({4\over \beta_0}\right)^{n+1}
\label{Esix}
\eea
where $\beta_0 = (33- 2N_f)/3$.
A discrete set of renormalon singularities
would show up as a set of finite-order poles in this plane
\beq
{r_k\over (y-y_k)^P}
\label{rPoles}
\eeq
The PA's in equation \eqref{PadeDef}
are clearly well suited to find the locations
$y_k$ and the residues $r_k$ of such poles.
In the case of a perturbative series dominated by a finite set of
$L$
renormalon singularities, a sufficiently high-order PA will be {\em exact}
\beq
[M/N](y) = \tilde S(y):\qquad {\rm for}~M+N > L_0
\label{PadeLimit}
\eeq
for some $L_0\propto L$.
Generically, in any case where the quantity analogous
to eq.~\eqref{epsDef}, $\tilde\epsilon_n\simeq 1/n^2$,
the error analogous to \eqref{Efour}  is given by
\beq
\tilde \delta_{[M/M]} \simeq - \,\frac{(M!)^2}{L^{2M}}
\label{Eeight}
\eeq
This prediction of very rapid convergence is confirmed in
Fig.~\ref{FigIII}b
\cite{SEK}
in the case of the QCD vacuum polarization D function evaluated in
the large $N_f$ limit \cite{Dfunction},
which has an infinite number of renormalon poles.

\begin{figure}[htb]
\begin{center}
\mbox{\epsfig{file=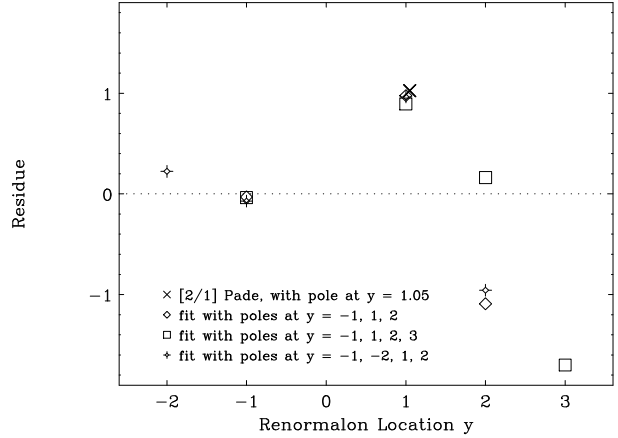,width=8truecm,angle=90}}
\end{center}
\caption{
The locations and residues of
poles \ in \ the \ [2/1] PA and \ in \ rational-function
fits to the Borel transform \ of
\ the first four terms \ in \ the perturbation series \ for
the Bjorken sum rule.
We note that the location of the lowest-lying infrared renormalon
pole is estimated accurately \ by Pad\'e Approximants in the Borel
plane, and that its residue is stable in the different fits.
\label{FigVII}
}
\end{figure}

    The power of the application of PA's in the Borel plane is
shown in Fig.~\ref{FigVII},
where we see that the [2/1] PA of the Borel-transformed
Bjorken series has a pole at
$y=1.05$\ .
The agreement with the expectation of a first infrared renormalon pole
at $y = 1$ is striking.  Fig.~\ref{FigVII} also shows other rational-function
approximations to the Borel-transformed Bjorken series.  We see evidence
that this is dominated by a strong infrared renormalon pole at $y = 1$,
that there may be a weak ultraviolet renormalon pole at $y = {-}1$ and
possibly another pole at $y = 2$.

    The ambiguity in the definition of the perturbative Bjorken series
associated with the $y = 1$ renormalon singularity corresponds to a
possible $1/Q^2$ correction of magnitude \cite{PBB}
\beq
\Delta\left(\Gamma^p_1 - \Gamma^n_1\right) = \pm \frac{|g_A|}{6}\,
0.98\,\pi\,\frac{\Lambda^2}{Q^2}
\label{bjrenamb}
\eeq
It is expected that the QCD correction to the Bjorken sum rule should
include a higher-twist correction of similar form with magnitude
\cite{HTrefs,BjSRalphas,BraunMoriond}
\beq
\Delta_{HT}\,\left(\Gamma_1^p - \Gamma_1^n\right)
= - \,\frac{0.02\pm0.01}{Q^2}
\label{bjht}
\eeq
The perturbative ambiguity in equation
\eqref{bjrenamb} is cancelled by a corresponding
ambiguity in the definition of the higher-twist contribution.  In the
next section we will treat equation \eqref{bjht} as a correction (with error)
to be applied to the perturbative QCD factor shown in Fig.~\ref{FigVI}.

    We have also compared PA's to the predictions of commensurate
scale relations within the framework of ref. \cite{CSR}.  The predictions
of the two approaches are numerically very similar, and we give
formal reasons in ref.~\cite{next}  why we believe that this should be so.
However, we shall not use commensurate scale relations in the data
analysis of the next section.

\subsection{Numerical analysis of the Bjorken sum rule}

     Table I shows the data on the integrals
$\Gamma_1^{p,n}$ currently available from experiments at CERN
and SLAC \cite{EMC},\hbox{\cite{PSFdata}-\cite{Roblin}}.
 We do not attempt to correct these numbers for any of the
effects discussed in section 1.4, such as the $Q^2$-dependence of
the asymmetry $A_1$, the extrapolation to low $x$, or the
transverse polarization asymmetry. We do not believe that any
of these effects will change any of the data outside their
quoted errors. We choose to evaluate the Bjorken sum rule at
$Q^2 = 3$ GeV$^2$, which requires rescaling all the data as
described in ref. \cite{PBB}.

\medskip
\def\pls{\phantom{{+}}}
\thicksize=1pt
\vbox{
\begin{center}
Table I \\
{\footnotesize
$\Gamma_1^{p,n,d}$ currently available from experiments at CERN
and SLAC.}
\end{center}
\medskip
\begintable
experiment | target | $\Gamma_1$                \crthick
E142       | $n$    | ${-}  0.045 \pm 0.009$    \cr
E143       | $p$    | $\pls 0.124 \pm 0.011$    \cr
E143       | $d$    | $\pls 0.041 \pm 0.005$    \cr
SMC        | $d$    | $\pls 0.023 \pm 0.025$    \cr
\ SMC ('94) \ | $d$    | $\pls 0.030 \pm 0.011$    \cr
SMC        | $p$    | $\pls 0.122 \pm 0.016$    \cr
EMC        | $p$    | $\pls 0.112 \pm 0.018$
\endtable
\begin{center}
{\footnotesize
All experimental data have been evolved to
$Q^2=3$ GeV$^2$.
}
\end{center}
} 

The following is the combined result
that we find for the Bjorken sum rule:
\beq
\Gamma_1^p(3 {\rm GeV}^2) - \Gamma_1^n(3 {\rm GeV}^2) = 0.164 \pm 0.011
\label{N}
\eeq
which is indicated by a vertical error bar
in the lower left corner of Fig.~\ref{FigVI}. Comparing this value
with the [2/2] PS estimate also shown there, we find
\beq
\alpha_s (3~{\rm GeV}^2) = 0.328^{{+}0.026}_{{-}0.037}
\label{asatQBj}
\eeq
which becomes
\beq
 \alpha_s(M_Z^2) = 0.119_{{-}0.005}^{{+}0.003}\,\, \pm \,\,\dots~\qquad,
\label{N+2}
\eeq
when we run $\alpha_s$ up to $M_Z^2$ using the three-loop
renormalization group equation. The errors quoted in equations
\eqref{asatQBj} and \eqref{N+2}
are purely experimental, and the second $\pm$ sign in
equation \eqref{N+2} indicates that further theoretical systematic errors
must be estimated. Those we have evaluated include that
associated with the renormalization scale dependence shown in
Fig.~\ref{FigV} ($\pm 0.002$), the difference between the [2/2] and
[2/1], [1/2] PS's ($\pm 0.002$), and the correction due to
the higher-twist estimate in equation
\eqref{bjht} ($ -0.003 \pm 0.002$),
whereas the error in the running of $\alpha_s$ is found to be
negligible. Combining these estimates with equation \eqref{N+2}, we find
\cite{PBB}
\beq
\alpha_s (M_Z^2) = 0.116_{-0.005}^{+0.003} \pm 0.003 ~,
\label{N+6}
\eeq
\begin{figure}[htb]
\begin{center}
\mbox{\epsfig{file=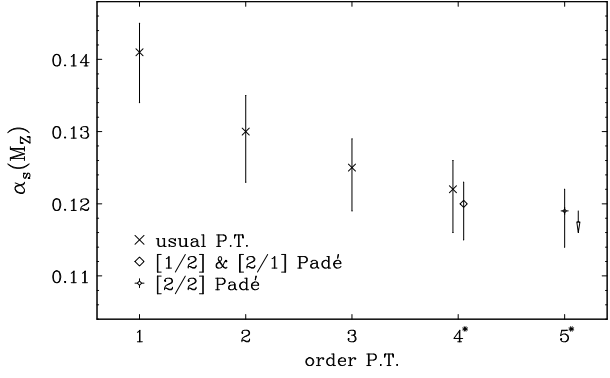,width=8truecm,angle=90}}
\end{center}
\caption{
Values of $\alpha_s(M_Z^2)$ obtained
using different orders \ of perturbation theory, \ compared \ with
\ our result \protect\eqref{N+2}, \ obtained using the [2/2] PS.
\ The size of the shift induced by higher-twist correction is
\protect\eqref{bjht}
indicated by a downward arrow to the right of the [2/2] point.
\label{FigVIII}
}
\end{figure}

The stability of this result is indicated in Fig.~\ref{FigVIII},
where we exhibit the values of $\alpha_s(M_Z^2)$ obtained
using different orders of perturbation theory, compared with
our result \eqref{N+6} obtained using the [2/2] PS. Also indicated is the
shift induced by the higher-twist correction, which lies within
our error bars.

\begin{figure}[htb]
\begin{center}
\mbox{\epsfig{file=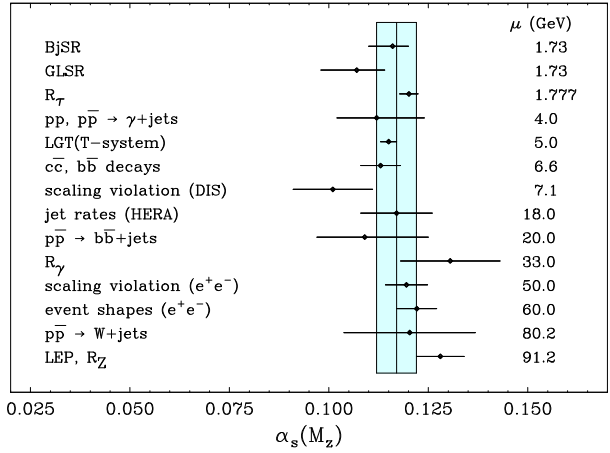,width=8truecm,angle=90}}
\end{center}
\caption{
\ \ Compilation \ of \ world data \ on
\ $\alpha_s$ \ from \ different sources
(adapted from  Ref.~\protect\cite{Schmelling}).
\label{FigIX}
}
\end{figure}

    As can by seen from the compilation in Fig.~\ref{FigIX}, our central
value for $\alpha_s(M_Z^2)$ is quite compatible with other
determinations and the world average, which is quoted to be
\cite{RPP,Bethke,Schmelling}
\beq
 \alpha_s(M_Z^2) = 0.117\pm 0.005
\label{alphasWorld}
\eeq
Indeed, the error quoted in equation \eqref{N+6}
is quite competitive
with the most precise determinations of $\alpha_s(M_Z^2)$ that
are available. Moreover, we note that plenty of precise new
data will soon be available from the SMC experiment at CERN,
the E154 and E155 experiments at SLAC, and the HERMES
experiment \cite{HERMES}
at DESY. In the longer run, experiments with
a polarized proton beam at HERA will provide valuable
information on the behaviour of $g_1$ at low $x$, as well as
on its $Q^2$-dependence at fixed $x$.

\subsection{Decomposition of the Nucleon Spin}

\begin{figure}[htb]
\begin{center}
\mbox{\epsfig{file=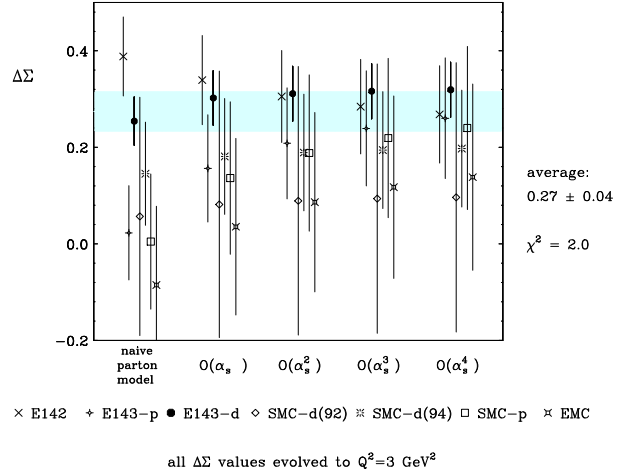,width=8truecm,angle=90}}
\end{center}
\caption{
The values of $\Delta\Sigma(Q^2{=}3\ \hbox{GeV}^2)$
extracted from each experiment, plotted as functions of the
increasing order of QCD perturbation theory
used in obtaining  $\Delta\Sigma$ from the data \
(from Ref.~\protect\cite{BjSRalphas}, updated with most recent data).
\label{FigX}
}
\end{figure}

So far, we have only discussed the combination $\Gamma_1^p -
\Gamma_1^n$ which enters in the Bjorken sum rule. The individual
$\Gamma_1^{p,n}$ can be used as in equation \eqref{GammaIDq}, though not
forgetting the perturbative QCD corrections in equation \eqref{ejCorr}, to
extract the individual $\Delta q$. As is seen in Fig.~\ref{FigX}, the
different experiments on both proton and neutron targets are
all highly consistent, {\it once the perturbative QCD corrections
are taken into account}. Some time ago, it appeared as if the
neutron data from E142 might be at variance with the other
data points. However, this is no longer the case if all the
higher-order corrections in equation \eqref{ejCorr} are taken into account,
and the latest evaluations \cite{Roblin}
of the E142 data indicate a different
preliminary value of $\Gamma_1^n$, as seen in Table I. Making a
global fit, we find
\vbox{
\bea
\Delta u &=& \phantom{-}0.82 \pm 0.03 \pm\ldots\nonumber \\
\Delta d &=& -0.44 \pm 0.03\pm\ldots\label{globalDqs} \\
\Delta s &=& -0.11 \pm 0.03\pm\ldots\nonumber
\eea
}
and
\beq
\Delta\Sigma =
\Delta u + \Delta d + \Delta s = 0.27\pm 0.04\pm\ldots
\label{globalDsigma}
\eeq
where the second $\pm$ sign indicates that further theoretical
and systematic errors remain to be assigned.
These include
higher-twist effects, errors in the extrapolation to low $x$
which is more complicated than for the nonsinglet combination
of structure functions appearing in the Bjorken integrand, the
possible $Q^2$-dependence of $A_1$, etc.. We believe that these
errors may combine to be comparable with the errors quoted in
equations \eqref{globalDqs},\eqref{globalDsigma},
but prefer not to quote definitive ranges for the
$\Delta q$ until all these errors are controlled as well as
those appearing in the Bjorken sum rule.

\begin{figure}[htb]
\begin{center}
\mbox{\epsfig{file=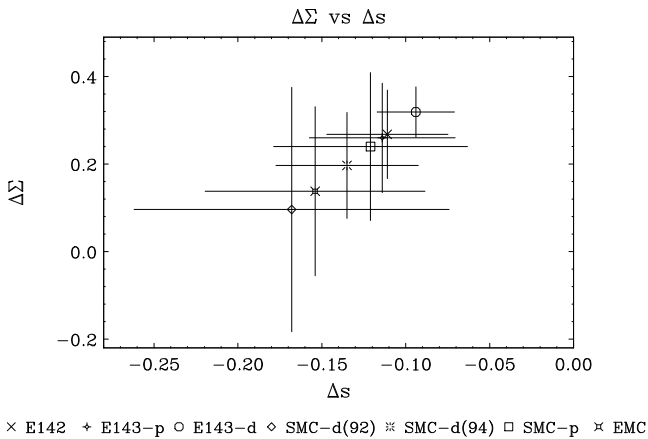,width=8.5truecm,angle=90}}
\end{center}
\caption{
\ \ The values \ of \ $\Delta\Sigma$ \ and \ $\Delta s$
extracted from each experiment, \ plotted \ against \ each other.
\ All \ data \ have been evolved \ to common $Q^2=3$ GeV$^2$.
The clear linear \ correlation
between \ \ $\Delta\Sigma$
\ \ and \ \ $\Delta s$ \ results \ \ from \ \ the \ \ linear \ \ relations
\protect\eqref{E12},\protect\eqref{E13},\protect\eqref{ejCorr}.
\label{FigXI}
}
\end{figure}

One may also get a feeling for the expected range of
$\Delta\Sigma$ and $\Delta s$ by plotting the
results for these two observables
extracted from each of the existing experiments,
as shown in Fig.~\ref{FigXI}.

\section{Outlook}

    In this talk we have concentrated on the phenomenological
analysis of the data on polarized structure functions presently
available. As we have seen, these tell a remarkably consistent
story, once higher-order QCD corrections are included. We have
not addressed in great detail here the
theoretical interpretation of the data,
nor their spin-offs in hadron physics and elsewhere,
nor possible future developments in this field. In
fact, these measurements provide valuable insights into important
issues in non-perturbative QCD, such as the \role\ of chiral
symmetry in nucleon structure \cite{BEK}, the axial anomaly and the $U(1)$
problem \cite{DeltagI}-\cite{DeltagIII},
and the relationship between current
and constituent quarks \cite{KaplanQ}-\cite{FHK}
which are provoking lively
theoretical debates (see Ref.~\cite{ConstPion} for a recent
application to the pion structure).

The polarized structure function data support
previous indications from the $\pi$-nucleon $\sigma$-term and
elsewhere that strange quarks in the nucleon wave function cannot
be neglected, with interesting implications for the analysis of
recent data from LEAR on $\phi$ production in proton-antiproton
annihilation \cite{EKKS}.
Among other spin-offs, we recall
that the axial-current matrix elements extracted from polarization
data determine scattering matrix elements for candidate dark
matter particles such as the lightest supersymmetric particle
\cite{SmallSpinI} and the axion \cite{BjSRalphas}.

In the future, we look
forward to the completion of the SMC programme and its possible
HMC successor at CERN, the E154 and E155 experiments at SLAC, data
from the HERMES experiment at HERA, the polarized proton programme
at RHIC, and possible polarized electrons and protons in the HERA
ring. The tasks of these experiments will include the determination
of the gluon contribution to the nucleon spin, the elucidation of
the $Q,^2$ dependence of $A_1$, and the low-$x$ behaviour of $g_1$.
These will continue to fuel activity in this interesting field for
the foreseeable future, which will lead us to a deeper understanding
of the nucleon, an object we thought we knew so well, but which
reveals a new face when it spins.

\begin{center}
{\large\bf Acknowledgements}
\end{center}
We thank Michelle Mazerand for her help in preparing the manuscript.
The research described in this talk
was supported in part by the Israel Science Foundation
administered by the Israel Academy of Sciences and Humanities,
and by
a Grant from the G.I.F., the
German-Israeli Foundation for Scientific Research and
Development.

\end{document}